\RequirePackage{fix-cm}
\documentclass[smallextended]{svjour3}       
\smartqed  
\usepackage{graphicx}

\usepackage{amsmath}
\usepackage{amsfonts}
\usepackage{amssymb}
\usepackage{mathrsfs} 
\usepackage[toc]{appendix}
\usepackage{pstricks}
\usepackage{pst-node,pst-3d,pst-text,pst-tree}
\usepackage{graphicx}
\usepackage{natbib}
\usepackage{color}
\usepackage{array}
\usepackage{algorithm2e}
\usepackage{graphicx}
\usepackage{arydshln}
\usepackage{enumerate}
\usepackage{empheq}
\usepackage{float}

\newtheorem{res}{Result}

\begin{document}

\title{The impracticalities of multiplicatively-closed codon models: a retreat to linear alternatives
\thanks{This research was supported by Australian Research Council (ARC) Discovery Grant DP150100088 to Barbara R. Holland and Jeremy G. Sumner and Australian Research Training Program scholarship to Julia A. Shore.}
}
%


\titlerunning{The impracticalities of multiplicatively-closed codon models}        

\author{Julia A. Shore         \and
        Jeremy G. Sumner  \and
        Barbara R. Holland
}


\institute{J. A. Shore \at
              Tel.: +61 3 6226 2439\\
              \email{julia.shore@utas.edu.au}           
           \and
           J. G. Sumner \at
              \email{jeremy.sumner@utas.edu.au}  
}

\date{Received: date / Accepted: date}

\maketitle

\begin{abstract}
A matrix Lie algebra is a linear space of matrices closed under the operation $ [A, B] = AB-BA $. The ``Lie closure'' of a set of matrices is the smallest matrix Lie algebra which contains the set. In the context of Markov chain theory, if a set of rate matrices form a Lie algebra, their corresponding Markov matrices are closed under matrix multiplication; this has been found to be a useful property in phylogenetics. Inspired by previous research involving Lie closures of DNA models, it was hypothesised that finding the Lie closure of a codon model could help to solve the problem of mis-estimation of the non-synonymous/synonymous rate ratio, $ \omega $. We propose two different methods of finding a linear space from a model: the first is the \emph{linear closure} which is the smallest linear space which contains the model, and the second is the \emph{linear version} which changes multiplicative constraints in the model to additive ones. For each of these linear spaces we then find the Lie closures of them. Under both methods, it was found that closed codon models would require thousands of parameters, and that any partial solution to this problem that was of a reasonable size violated stochasticity. Investigation of toy models indicated that finding the Lie closure of matrix linear spaces which deviated only slightly from a simple model resulted in a Lie closure that was close to having the maximum number of parameters possible. Given that Lie closures are not practical, we propose further consideration of the two variants of linearly closed models.
\end{abstract}

\section{Introduction}
\label{intro}

It is of interest to evolutionary biologists to determine what biological and chemical mechanics contribute to the evolution of genomes. By making comparisons between rates of substitutions in amino acids, changes in the functionality of a genome can be studied. Alternatively, a comparison of rates of substitutions of DNA nucleotides allows observation of the underlying random processes of genome evolution. Both the studies of functionality and underlying processes are of interest to biologists as their simultaneous consideration allows for more accurate modelling of evolutionary data. The analysis of codons (triplets of DNA, each of which codes for an amino acid) allows one to analyse factors of both DNA rates of change and amino acid rates of change at once. 

A typical characteristic of codon models is inclusion of a non-synonymous/ synonymous relative rate, $ \omega $, as a model parameter. A synonymous mutation is one between two codons that code for the same amino acid and hence the functionality of the gene does not change under such a mutation. For example codons AAA and AAG both code for the amino acid lysine so a mutation from AAA to AAG is synonymous. A non-synonymous mutation is one where the amino acid does change. For example AAA and AAC code for amino acids lysine and asparagine respectively, so a mutation between these codons would change the functionality of the gene. The parameter $ \omega $ is the relative rate of non-synonymous to synonymous mutations. Values of $\omega$ are characterised as follows:

\[\begin{cases}
\omega <1 & \text{ purifying selection} \\
\omega =1 & \text{ neutral evolution} \\
\omega >1 & \text{ positive selection}.
\end{cases}\]
\noindent in practice, purifying selection (that is, synonymous mutations happen more frequently than non-synonymous mutations) is most often observed. The observation of neutral evolution can indicate that the gene in question is unimportant as a mutation that changes the protein produced is equally likely to one that does not. There are cases of positive selection occurring, this is most often observed in viruses \citep{bennett2006molecular,shen2009diversifying,yang1998likelihood}. 

As there are four DNA nucleotides, there are 64 ($ = 4 \times 4 \times 4 $) codons which may lead one to think that such a model would be computationally expensive to use when it is considered that the maximum number of free rate parameters for a codon model would be $ 4032 $ ($ = 64 \times 64 - 64 $). Despite this, codon models currently in use can simultaneously model aspects of both functionality of a gene and underlying DNA process with as few as two free parameters. The branch-site codon models described in \citet{yang1997}, for example, can have as few as a single parameter which is the proportion of codon sites whose $ \omega $ values fall in a specified range. The Muse-Gaut codon model \citep{muse1994likelihood} takes into account $ \omega $ and the frequency of DNA nucleotides; a total of four free parameters. 


A common way to represent models of evolution is with matrices where an off diagonal $ (i,j)^{th} $ entry of rate matrix $Q$ represents the rate in which state $j$ changes to state $i$ and diagonal entries are chosen to give $Q$ zero column sum. Note here that this differs to the similar common convention of an off-diagonal $ (i,j)^{th} $ entry of such a matrix to represent the rate of state $i$ changing to state $j$ and hence off-diagonal entries of the matrix are chosen to give zero row sum. Mathematically, a model which contains free parameters can be represented by a \emph{set} which here we generically denote as $ \mathcal{Q} $ so $ Q \in \mathcal{Q} $. In this paper, we assume that the matrix set $\mathcal{Q}$ is always determined by polynomial constraints (on the matrix entries). Further we note that in most cases constraints are homogeneous. In these cases, for any $Q \in \mathcal{Q}$, scalar multiples $\lambda Q$ are also in $\mathcal{Q}$. For any given rate matrix $Q$ it is possible to generate a corresponding transition matrix $M$ where an off-diagonal $ (i,j)^{th} $ entry of $M$ represents the probability of state $j$ changing to state $i$ in a given time period and the diagonal entries are chosen to give a unit column sum. This is done through the exponential map: $ M = e^{Qt} $ where $t$ represents time elapsed. It should be noted here that the $n \times n$ zero matrix (where $n$ is the number of states) is always assumed to be contained in $\mathcal{Q}$ and, as a consequence of taking the matrix exponential of the zero matrix, the $n \times n$ identity matrix is always contained in the corresponding set of Markov matrices.


In the case of DNA models, research by \citet{sumner2011} found practical merit in having a set of Markov matrices which are closed under matrix multiplication. If there are two Markov matrices, $ M_{1} $ and $ M_{2} $, acting on different segments on the same branch of a phylogenetic tree, in order to find the overall process, $ \hat{M} $, for that branch it is required to multiply $ M_{1} $ and $ M_{2} $ together. Therefore, if a set of Markov matrices, $ \mathcal{M} $, are closed under matrix multiplication and $ M_{1}, M_{2} \in \mathcal{M} $, then $ \hat{M} $ in this scenario would also belong to $ \mathcal{M} $. For a set of rate matrices $\mathcal{Q}$ and its corresponding set of transition matrices $\mathcal{M}$, it has been shown that $\mathcal{M}$ is closed under matrix multiplication if and only if $\mathcal{Q}$ forms a Lie algebra \citep{sumner2017multiplicative, sumner2011}. Therefore, demanding that $ \mathcal{Q} $ forms a Lie algebra will ensure $ \hat{M} \in \mathcal{M} $. Further studies by \citet{sumner2011} found that the general time reversible model \citep{tavare1986} (from now referred to as GTR) does not have this property, i.e. if $ M_{1} $ and $ M_{2} $ are of GTR form then it is not always the case that $ \hat{M} $ is also of GTR form.

\citet{woodhams2017exploring} conducted similar research on codon models. Their simulations involved selecting a DNA model, generating two sets of parameters from these models to produce two distinct codon Markov matrices (in Section \ref{sec:1} we describe this process in detail). They then demonstrated that if these two phylogenetic processes on the two branches of a two-taxon phylogenetic tree have the same underlying value of $ \omega $ then the average process over the tree is estimated, the resulting process does not necessarily have the same value of $ \omega $. This result was consistent for different initial DNA models selected from both time-reversible models and Lie-Markov models (LMM) \citep{sumner2011,fernandez2015lie}. It seems sensible given previous research on the mis-estimation of substitution probabilities in DNA models to assume that a codon model which forms a Lie algebra would be less prone to mis-estimation of $ \omega $. It is currently an open problem to construct such a model. The purpose of this paper is to explore the inherent obstructions in doing so. 

In this paper, we explore two methods of finding a Lie algebra to represent a codon model. Both these methods compute the smallest Lie algebra containing a given linear space; the methods differ by how they generate a linear space from a codon model. The first method is to find the ``linear closure'' which is the smallest linear space containing the model (in this case the Lie algebra generated by using the method is the smallest Lie algebra which contains a model). The second method is to find the ``linear version'' which is found by changing the operations in a codon model's formulation in such a way that the resulting space is linear. We found that both methods produce inherit difficulties concerning the large number of parameters in the resulting codon model. The method of finding the smallest Lie algebra which contains a model has the additional difficulty of there not being an unambiguous definition of the $ \omega $ parameter.

As it has been shown that multiplicative closure of a set of Markov matrices occurs if and only if the rate matrices form a Lie algebra \citep{sumner2017multiplicative}, our results show that there is no practical way to have a multiplicatively closed codon model. This tells us that there is a fundamental conflict between codon models, the genetic code, and multiplicative closure. 

However, although our attempt to find a codon model which is a Lie algebra gave impractical results, the linear spaces generated during the process of this analysis (the linear closures and linear versions) potentially offer a partial solution to the initial problem. These linear codon models do not have the problems associated with codon models that are Lie algebras: they have a sensible number of parameters and, unlike other partial solutions we discuss, have positive substitution rates. Linear closures and linear versions of codon models closely resemble the initial codon models from which they are generated and in the context of multiplicative closure it has been shown that in the DNA case a linear model mis-estimates parameters less than a non-linear model \citep{bodie2011effect}.


Addressing the issue of multiplicative closure of Markov models is not the first application of Lie theory in the field of genetics. There have been several studies which find Lie algebras and similar structures which resemble different aspects of the genetic code, such as the relationship between codons and amino acids \citep{hornos1993algebraic,bashford1998supersymmetric,bashford2000genetic,sanchez2006novel}. These works, however, bear little relevance to our study because we are aiming to use the structure of Lie algebras to build a Markov model which represents mutation rates between codons, not to represent the genetic code itself with an algebraic structure.


In Section \ref{sec:1} of this paper we define commonly used codon models and explore their structure. Next, in Section \ref{sec:2}, the mathematical tools used to find linear spaces to represent codon models are defined and examples are given for finding linear spaces associated with sets of matrices, some of which are Markov chains which represent DNA nucleotide substitutions. Also in Section \ref{sec:2}, the procedure of finding a Lie algebra from a linear space is defined and examples are given. Our analysis and its results are then discussed. In Section \ref{sec:3} we use a toy model to illustrate why the size of the Lie algebras of codon models tend to be so large. Finally in Section \ref{sec:4}, the results of this study are summarised and possible further research topics are explored.

\section{Defining the codon model} 
\label{sec:1}

The Muse-Gaut codon model (from now referred to as MG) \citep{muse1994likelihood} defines the rate of change from codon $J = (j_{1},j_{2}, j_{3}) $ to codon $I = (i_{1},i_{2},i_{3})$  as follows:

\[ Q_{IJ} =  \begin{cases} 
      \pi_{i_{k}} & \text{synonymous} \\
      \omega \pi_{i_{k}} & \text{non-synonymous} \\
      0 & \text{multiple nucleotide substitutions needed}
   \end{cases}
\]

\noindent where $I \not= J$ and $k$ is the codon position that is undergoing a mutation and $ \pi_{i_{k}} $ is the frequency of nucleotide $ i_{k} $. The diagonal entries of $ Q_{IJ} $ are chosen to give zero column sum. 

We will be studying MG ``style'' codon models, which are based on the original MG model, as described presently.

Following the derivation given in \citet{woodhams2017exploring}, we first consider DNA substitution matrices, $M_{1}$, $M_{2}$ and $M_{3}$, whose entries give the probabilities of DNA substitutions at codon positions $1$, $2$ and $3$ respectively. With the assumption of independence of mutations at codon sites, it follows that the probability of transition from the triplet $J$ to $I$ is given by the product

\[ \text{prob}(j_{1}j_{2}j_{3} \rightarrow i_{1}i_{2}i_{3}) = M_{1}(i_{1},j_{1})M_{2}(i_{2},j_{2})M_{3}(i_{3}j_{3}). \]

In the construction of this set of models, we use $ \otimes $ to signify the Kronecker product of matrices. As an example of this operation, consider the two matrices:

\[ A = \left( \begin{array}{cc} \phantom{-}3 & \phantom{-}1 \\ \phantom{-}0 & -2 \end{array} \right), B = \left( \begin{array}{cc} \phantom{-}4 & \phantom{-}2 \\ -1 & \phantom{-}1 \end{array} \right). \]

\noindent The Kronecker product of $A$ and $B$ is as follows:

\[ A \otimes B = \left( \begin{array}{cc} \phantom{-}3B & \phantom{-}1B \\ \phantom{-}0B & -2B \\\end{array} \right) = \left( \begin{array}{cccc} \phantom{-}12 & \phantom{-}6 & \phantom{-}4 & \phantom{-}2 \\ -3 & \phantom{-}3 & -1 & \phantom{-}1 \\ \phantom{-}0 & \phantom{-}0 & -8 & -4 \\ \phantom{-}0 & \phantom{-}0 & \phantom{-}2 & -2 \\ \end{array} \right). \]

The Kronecker product can then be used to represent the transition probabilities from codon $J$ to $I$. Specifically, the matrix $ M_{triplet} = M_{1} \otimes M_{2} \otimes M_{3} $ is $ 64 \times 64 $ and contains the transition probabilities of codons based on the underlying DNA processes. 

Now assuming a continuous time Markov chain, we may recover a rate matrix, $Q$ from a transition matrix, $M(t)$, by taking the derivative of $M(t)$ and evaluate it at $t=0$. By applying this operation to $M_{triplet}$ we are given the result:

\begin{equation} Q_{triplet} = Q_{1} \otimes I \otimes I + I \otimes Q_{2} \otimes I + I \otimes I \otimes Q_{3}, \label{Qtrip} \end{equation}

\noindent where $I$ is the $ 4 \times 4 $ identity matrix. Here, the $Q_{k}$ describe the underlying DNA rates of change processes of different codon positions: $Q_{1}$ for position 1, $Q_{2}$ for position 2 and $Q_{3}$ for position 3. Typically one takes the constraint that $Q_{1} = Q_{2} = Q_{3}$, i.e. the underlying DNA rate of change process is the same at all three codon positions. We recall that an assumption almost always made in codon models is that the rate of multiple substitutions is zero, \emph{e.g.} the rate of $ AAA \rightarrow AGG =0 $ as two DNA mutations are required. This property is automatically present in the $ Q_{triplet} $ matrix as given: the zero matrix entries of the matrix $I$ ensure that entries of $Q_{triplet}$ are zero for multiple substitutions.

We define a $ 64 \times 64 $ matrix $G$ to contain information about the rates of change of codons with respect to their amino acid counterparts. It  contains $ \omega $ for non-synonymous substitutions, $1$ for synonymous substitutions and $0$ for prohibited substitutions (to and from stop codons). We then have:

\begin{definition}

\textit{MG-style codon models} are given by

\[ Q_{codon} = Q_{triplet} \circ G, \]

\noindent where the $ \circ $ operation first computes the element-wise product of the two matrices and then resets the diagonal entries to ensure zero column sum.

\end{definition}

In the original MG paper, the F81 model \citep{felsenstein1981} was assumed for the underlying DNA rate substitution process. However, any DNA rate process can be used to produce a codon model in the above formulation. In the following analysis, we illustrate our discussion with the Kimura 2 parameter model (from now referred to as K2ST) \citep{kimura1980simple} and the Jukes Cantor model (from now referred to as JC) \citep{jukes1969} as our underlying DNA rate processes. An MG process with an underlying model of JC is denoted as JC-MG, and similarly, an MG process with K2ST as the underlying model is denoted as K2ST-MG.

As noted above, for any DNA or codon rate matrix, we can find the corresponding transition matrix (that is, a matrix whose entries are the probability of the change of state for a given time period) by taking the matrix exponential of that rate matrix multiplied by time elapsed as per Markov chain standard practice.

\section{Finding the linear and Lie closures of a model} 
\label{sec:2}
In this section, both the general process for finding the Lie closure of an arbitrary set of matrices and the specific results of finding the Lie closures for models JC-MG and K2ST-MG are discussed.

For a given codon model, we consider two methods for finding a linear space associated to it. Firstly, we look at the smallest linear space that contains the model. This is a fairly straightforward process as we presently illustrate. We define Mat$_{n \times m}(\mathbb{R}) $ to be the set of all $ n \times m $ matrices with real entries. Note that  Mat$_{n \times m}(\mathbb{R}) $ forms a linear space under matrix addition.

\begin{definition}
The \emph{linear closure} of a set of $ n \times m $ matrices with polynomial constraints on the matrix entries, $ \mathcal{A} $, is the intersection of all linear sub-spaces of Mat$_{n \times m}(\mathbb{R}) $ which contain $ \mathcal{A} $. More simply, this can be described as the smallest linear space which contains $ \mathcal{A} $.
\end{definition}

For the examples we consider in this paper, to obtain the linear closure of $\mathcal{A}\subseteq \text{Mat}_{n\times n}(\mathbb{R})$ it is sufficient to take the set, $\mathcal{X}$, of polynomial constraints on the matrix entries of members of $\mathcal{A}$, and remove the non-linear polynomials in $\mathcal{X}$. Caution is required however, since in general there exists cases where this process gives the incorrect answer for the linear closure. Characterising the cases where our procedure will produce the linear closure of a matrix set is, in general, a difficult topic that we do not address in this paper.





\begin{example}
\label{firstLC}

As an example of the process of finding the linear closure of a matrix set, we find the linear closure for the following set of matrices defined using two parameters

\[ \mathcal{A} =  \left\{ \left( \begin{array}{cc} x & xy \\ y & y \end{array} \right): x,y \in \mathbb{R} \right\}.  \]

Equivalently we may define this set in terms of constraints on the matrix entries: $ \mathcal{X} = \{ A_{11}A_{22} = A_{12} , A_{21} = A_{22} \} $. Applying our method, the first constraint is non-linear and is discarded. The second constraint is linear so this remains as a constraint. Our method therefore gives the set,

\begin{align*} \mathcal{A}' =& \left\{ A \in \text{Mat}_{2 \times 2}: A_{21} = A_{22} \right\} \\ =&  \left\{ \left( \begin{array}{cc} x & z \\ y & y \end{array} \right): x,y,z \in \mathbb{R} \right\} \\ =& \text{span}_{\mathbb{R}}  \left\{
\left( \begin{array}{cc} 1 & 0 \\ 0 & 0 \end{array} \right),
\left( \begin{array}{cc} 0 & 1 \\ 0 & 0 \end{array} \right),
\left( \begin{array}{cc} 0 & 0 \\ 1 & 1 \end{array} \right) \right\} 
\end{align*}

\noindent which is a three dimensional linear space.

As stated above, there are cases where our method of removing all non-linear constraints does not result in the linear closure. To show that $\mathcal{A}'$ is in fact the linear closure of $\mathcal{A}$, we note that the linearly independent matrices 

\[ 
\left( \begin{array}{cc} 1 & 1 \\ 1 & 1 \end{array} \right),
\left( \begin{array}{cc} 1 & 2 \\ 2 & 2 \end{array} \right) \text{ and }
\left( \begin{array}{cc} 2 & 2 \\ 1 & 1 \end{array} \right) \]

\noindent are all contained in $ \mathcal{A} $, which tells us that the linear closure must be dimension $3$ or more. The dimension of $ \mathcal{A}' $ is $3$ and it is clear that $ \mathcal{A} \subseteq \mathcal{A}' $, so the linear closure of $\mathcal{A}$ must have dimension $3$ or less. Hence we conclude that $\mathcal{A}'$ is indeed the linear closure of $\mathcal{A}$.
\end{example}

An alternative process to this is that instead of considering the linear closure of a space, we explore the ``linear version'' as described presently.

\begin{definition}
For a set $\mathcal{A} \subseteq $Mat$_{ n \times m}(\mathbb{R}) $ of matrices defined using polynomial constraints, a fixed non-zero $n \times m$ matrix $B \in \mathcal{A}$, and smooth paths $A(t) \in \mathcal{A}$ which start at $B$, i.e. $A(0)=B$, we define $ T_{B}(\mathcal{A})$ to be the \textit{tangent space of $ \mathcal{A} $ at $B$}. Specifically, this is the span of derivatives of these paths evaluated at zero: $ T_{B}(\mathcal{A}):= \text{span}_{\mathbb{R}}\{A'(0): A(t) \in \mathcal{A} \text{  } \forall t, A(0)=B$\}.
\end{definition}

In the language of algebraic geometry, we note that the $ T_{B}(\mathcal{A}) $ is in fact the linear space associated to the linear variety through $B$. To clarify, we explore this concept in Example \ref{linVarEx}.

We also note there are some cases where the only smooth path in a matrix set from some matrix $B$ is trivial i.e. $ A(t) = B $  $\forall t$. The derivative of such a path is $0$. 


\begin{example}
\label{linVarEx}
Consider the unit circle, $x^2 + y^2 = 1$, on $ \mathbb{R}^2 $. The tangent line on the circle at  $(1,0)$ gives us the set of points $ \mathcal{M} = \{ (1,a): a \in \mathbb{R} \} $ which all sit on the line $x=1$. We note that $ \mathcal{M} $ does not form a linear space as it in particular does not contain $(0,0)$. 

Now we consider a path on the unit circle $x(t)^2 + y(t)^2 = 1$ where $x(0)=1$ and $y(t)=0$. Differentiating with respect to $t$ gives us $ 2(x(t)x'(t) + y(t)y'(t)) = 0 $. When we let $t=0$, the condition becomes $2x'(0) = 0$. Hence the tangent space is $ \mathcal{N} = \{ (0,a): a \in \mathbb{R} \} $. Unlike $\mathcal{M}$, $\mathcal{N}$ is a linear space.

In this example, $\mathcal{M}$ is the linear variety through $(1,0)$ whereas $\mathcal{N}$ is the tangent space at $(1,0)$. Under our terminology, the point $(1,0)$ is not in the tangent space at $(1,0)$.
\end{example}

\begin{definition}
Given a set, $ \mathcal{A} $, of $ n \times m $ matrices defined using polynomial constraints, whose terms have degree $1$ or greater and a fixed $ n \times m $ matrix, $B$, the \emph{linear version} of $ \mathcal{A} $ at $B$ is $ T_{B}(\mathcal{A}) $.
\end{definition}

\begin{lemma}
Given a set, $ \mathcal{A} $, of $ n \times m $ matrices with polynomial constraints of positive degree on the matrix entries:  \label{kitten}

i) For any $B\in \mathcal{A}$, $ T_{B}(\mathcal{A}) $ is a subspace of the linear closure of $\mathcal{A}$.

ii) In the situation that $\mathcal{A}$ is defined using only homogeneous polynomial constraints of positive degree, $ T_{0}(\mathcal{A}) $ is equal to the linear closure of $\mathcal{A}$.
\end{lemma}

\begin{proof}
\textit{i)} By the definition of the derivative, members of the linear version of $ \mathcal{A} $ at $B$ are limits of members of the linear closure of $ \mathcal{A} $ and real linear spaces are closed under limits. \\
\textit{ii)} Setting $B=0$, part (i) tells us that $T_0(\mathcal{A})$ is a subspace of the linear closure of $\mathcal{A}$.

Now consider $C\in \mathcal{A}$. Then, by the homogeneity assumption, each $A(t):=tC $ is a smooth path in $\mathcal{A}$. This immediately implies $C=A'(0)\in T_0(\mathcal{A})$ and hence $T_0(\mathcal{A})$ contains $\mathcal{A}$. But then, since $T_0(\mathcal{A})$ is a subspace of Mat$_{n\times m}(\mathbb{R})$, it must contain the linear closure of $\mathcal{A}$.

The result follows.
\qed
\end{proof}

We note that in the situation that $\mathcal{A}$ is already a linear space, the linear version of $\mathcal{A}$ at (any) $B \in \mathcal{A}$ and linear closure of $ \mathcal{A} $ are both equal to $\mathcal{A}$.

\begin{lemma}
To find a linear version of a set of matrices with polynomial constraints where $B$ has unit entries everywhere, any polynomial constraints of degree $ \geq 2 $ on the original matrix set are replaced by changing each multiplication operation to an addition operation and any non-zero constant terms in the polynomial constraints are replaced by $0$. \label{burrich}
\end{lemma}

\begin{proof}
For a matrix set $ \mathcal{A} $ with polynomial constraints $\mathcal{X}$ on the matrix entries, consider arbitrary $ C \in \mathcal{A} $ and $ f \in \mathcal{X} $: 

\[ f(C) = a_{1}c_{11}^{k_{11;1}}c_{12}^{k_{12;1}}...c_{nm}^{k_{nm;1}} + ... + a_{p}c_{11}^{k_{11;p}} c_{12}^{k_{12;p}}...c_{nm}^{k_{nm;p}} = 0. \]

\noindent where $ a_{i} \in \mathbb{R} $, $ c_{ij}  $ are the matrix entries of $C$ and $ k_{ij;l} \in \mathbb{N} \cup \{ 0 \} $. To find the tangent space of $ \mathcal{A} $ at $B$, we first consider paths in $ C(t) \in \mathcal{A} $ with matrix entries $ c_{ij}(t) $ where $ c_{ij}(0) = 1 $ (\emph{i.e.} $ C(0) = B $). Therefore $ f \in \mathcal{X} $ becomes

\begin{align*} f(C(t)) =& a_{1}c_{11}(t)^{k_{11;1}}c_{12}(t)^{k_{12;1}}...c_{nm}(t)^{k_{nm;1}} + ... \\
&+ a_{p}c_{11}(t)^{k_{11;p}} c_{12}(t)^{k_{12;p}}...c_{nm}(t)^{k_{nm;p}} \\
=& 0 .
\end{align*}

\noindent To find the tangent space, we differentiate the constraints on $ \mathcal{A} $ with respect to $t$ and set $ t=0 $ (as $ C'(0) \in $ tangent space at $B$):

\begin{align*} 
\left. \frac{d}{dt}\right|_{t=0 }&  (a_{1}c_{11}(t)^{k_{11;1}}c_{12}(t)^{k_{12;1}}...c_{nm}(t)^{k_{nm;1}} + ... \\
& + a_{p}c_{11}(t)^{k_{11;p}} c_{12}(t)^{k_{12;p}}...c_{nm}(t)^{k_{nm;p}}) \\
 = \left. \frac{d}{dt} \right|_{t=0} & 0 =0, 
\end{align*}

\noindent then using the chain rule and product rule, we differentiate the first term only:

\begin{align*} & \left. \frac{d}{dt} \right|_{t=0}  a_{1}c_{11}(t)^{k_{11;1}}c_{12}(t)^{k_{12;1}}...c_{nm}(t)^{k_{nm;1}} \\ =& a_{1}( k_{11;1} c_{11}'(0) c_{11}(0)^{k_{11;1}-1} c_{12}(0)^{k_{12;1}}...c_{nm}(0)^{k_{nm;1}}  \\
 &+ k_{12;1} c_{12}'(0) c_{11}(0)^{k_{11;1}} c_{12}(0)^{k_{12;1}-1}...c_{nm}(0)^{k_{nm;1}} + ... \\
 &+ k_{nm;1} c_{nm}'(0) c_{11}(0)^{k_{11;1}} c_{12}(0)^{k_{12;1}}...c_{nm}(0)^{k_{nm;1}-1}) \\
 =& a_{1} ( k_{11;1} c_{11}'(0) + k_{12;1}c_{12}'(0) + ... + k_{nm;1}c_{nm}'(0) ) \end{align*}

\noindent as $ c_{ij}(0) = 1$ $ \forall i,j $. Note that this expression would contain no constant terms as the process of differentiation sends all constant terms to zero. A similar procedure can be used for the other terms of $f$ so that the tangent of the constraint at $B$ is

\begin{align*}
  a_{1}({k_{11;1}}c_{11}'(0) + {k_{12;1}}c_{12}'(0) +...+ {k_{nm;1}}c_{nm}'(0)) + ...  & \\ +  a_{p}({k_{11;p}}c_{11}'(0) + {k_{12;p}}c_{12}'(0) +... + {k_{nm;p}}c_{nm}'(0) )  & = 0
\end{align*}
\noindent which is equivalent to changing every multiplication operation between $ c_{ij} $ entries with addition. As $f$ was an arbitrary element of $\mathcal{X}$, we can say that this applies to all $ f \in \mathcal{X} $.
 \qed
\end{proof}

In the case of sets of DNA rate matrices, $ \mathcal{Q} $, whose constraints on matrix entries are polynomial we take $B$ (the point at which we take the tangents) as $B=J_{DNA}$ which we define as the JC matrix with DNA rate of change equal to $1$ i.e. the $4 \times 4$ matrix which has unit value off-diagonal entries and whose diagonal entries ensure zero column sum.

For MG-style codon models, we set $B = J_{codon}$ where
\[ J_{codon}(i,j) = \begin{cases}
0 & i \not=j \text{ and is a STOP codon row or column} \\
0 & i \not= j \text{ and multiple DNA substitutions are required} \\
1 & i \not= j \text{ and one DNA substitution is required} \\
a_{i} & i=j \text{ and } a_{i} = -\sum_{k=1, k \not=i}^{64} B_{ik}, \\
\end{cases}
\]

\noindent i.e. $J_{codon}$ has unit entries anywhere where we expect to see a non-zero rate of change in a codon model. Here we note that for matrix $G'$ being $G$ with $\omega = 1$, we have

\[  J_{codon} =(J_{DNA} \otimes I \otimes I + I \otimes J_{DNA} \otimes I + I \otimes I \otimes J_{DNA} ) \circ G'.  \]

When it is clear from the context, both $J_{DNA}$ and $J_{codon}$ will be abbreviated to $J$.

Note that whenever scale multiples of the matrix $B$ lie within the model under consideration (which is certainly the case for $J_{codon}$ in the MG-style models considered here), the matrix $B$ is guaranteed to also be in the linear version of the model: $B$ occurs as the tangent vector to smooth paths on the line that goes from $0$ to $B$. One obvious such path is $A(t):=(1+t)B$, which gives $A(0)=B=A'(0)$. We also note that finding the tangent space of a matrix space $\mathcal{A}$ at $0$ as opposed to $B$ would give a different space as a result. In particular: Lemma \ref{kitten} states that in the case of homogeneous constraints, $ T_{0}(\mathcal{A}) $ and the linear closure are identical.

To find the linear version of $ \mathcal{Q} $, we treat constraints on off diagonal entries the same way as they are treated in Lemma \ref{burrich} ($c_{ij}(0) = 1$) and treat diagonal entries to be the entry required to ensure zero column sum.

\begin{example}
\label{firstLV}
Using the same matrix set from Example \ref{firstLC}, we take the linear version at $B$ being the $ 2 \times 2 $ matrix with unit entries. The constraint of $ A_{11}A_{22} = A_{12} $ is changed to $ A_{11} + A_{22} = A_{12} $ so that the linear version of the set would be 

\[ \left\{ \left( \begin{array}{cc} x & xy \\ y & y \end{array} \right): x,y \in \mathbb{R} \right\} \rightarrow
\left\{ \left( \begin{array}{cc} x & x+y \\ y & y \end{array} \right): x,y \in \mathbb{R} \right\} = \text{span}_{\mathbb{R}} \left\{ \left( \begin{array}{cc}  1 & 1 \\ 0 & 0 \\ \end{array} \right), \left( \begin{array}{cc} 0 & 1 \\ 1 & 1 \\ \end{array} \right) \right\} \]

\noindent which has two free parameters, hence is a two dimensional matrix linear space. Note that the number of free parameters in the linear version is the same as the original.
\end{example}

\begin{example}
\label{HKYlclv}

As a more practical example of these processes, we examine a variation of the HKY model \citep{hasegawa1985} by adding the constraints of $ \pi_{A} = \pi_{G} $ and $ \pi_{C} = \pi_{T} $. The matrix representation of this model has three free parameters:

\[ \mathcal{Q}_{HKY'} = 
\left\{ \left( 
\begin{array}{cccc}
* & \pi_{A} \kappa & \pi_{A} & \pi_{A} \\ 
\pi_{A} \kappa & * & \pi_{A} & \pi_{A} \\ 
\pi_{C} & \pi_{C} & * & \pi_{C} \kappa \\ 
\pi_{C} & \pi_{C} & \pi_{C} \kappa & * \\ \end{array} \right): \pi_{A}, \pi_{C}, \kappa \in \mathbb{R} \right\} \]

\noindent where * denotes that the entry is chosen to give zero column sum.

To calculate the linear closure of this model, linear constraints such as $ A_{14} = A_{24} $ are kept but non-linear constraints such as $ A_{12}A_{41} = A_{13}A_{43} $ are discarded. Thus the linear closure of this model is
\[ \left\{ \left( \begin{array}{cccc} * & \gamma & \alpha & \alpha \\ \gamma & * & \alpha & \alpha \\ \beta & \beta & * & \delta \\ \beta & \beta & \delta & * \\ \end{array} \right): \alpha, \beta, \gamma, \delta \in \mathbb{R} \right\} \]

\noindent which is Model 4.4b (a four dimensional matrix linear space) in the Lie-Markov model (LMM) hierarchy described in \citet{fernandez2015lie}. This is different to the linear version, with respect to $B=J$ as defined above, which would change the non-linear constraints such as $ A_{12}A_{41} = A_{13}A_{43} $ to linear constraints such as $ A_{12} + A_{41} = A_{13} + A_{43} $ so that the linear version of the model is

\[ \left\{ \left( \begin{array}{cccc} * & \alpha + \kappa & \alpha & \alpha \\ \alpha  + \kappa & * & \alpha & \alpha \\ \beta & \beta & * & \beta + \kappa \\ \beta & \beta & \beta + \kappa & * \\ \end{array} \right): \alpha, \beta, \kappa \in \mathbb{R}  \right\} . \]

\noindent This is Model 3.4 (a three dimensional matrix linear space) in the LMM hierarchy \citep{fernandez2015lie}. 
\end{example}

One might notice that in these examples of finding linear versions of DNA models, instances where parameters are multiplied together are changed to the parameters being added together. In the case of the last example, we can imagine that if transitions were less likely to occur than transversions we would expect $ \kappa <1 $ for the model but $ \kappa <0 $ for the linear version (this is assuming a DNA ordering of AGCT). Having negative additive parameters in a model can potentially bring about negative substitution rates if not handled carefully. There are many examples, however, of this issue being overcome computationally by using appropriate constraints on the parameters. For the Lie-Markov models, a sensible approach is described in \citet{woodhams2015new}. For this example, the constraints required to maintain stochasticity are $ \alpha>-\kappa $ and $ \beta>-\kappa $. 

In our analysis, we will be using both Lie closures (which are the Lie closures of the linear closures) of codon models and Lie closures of linear versions of codon models.


\begin{definition}
	A \emph{Lie algebra}, $ \mathcal{L} $, is a linear space over a field, $ \mathbb{F} $, with an additional operation of the Lie bracket $ [\cdot,\cdot]: \mathcal{L} \times \mathcal{L} \rightarrow \mathcal{L} $, which, for $ x,y,z \in \mathcal{L} $ and $ \lambda \in \mathbb{F} $ satisfies:
	\begin{enumerate}
		\item[(i)] $ [x,y] = -[y,x], $
		\item[(ii)] $ [ \lambda x,y ] = \lambda[x,y], $
		\item[(iii)] $ [x, [y,z] ] + [ y, [z,x] ] + [z, [x,y] ] = 0. $
	\end{enumerate}
\end{definition}

In matrices, we define the Lie bracket operation as $ [A,B] = AB-BA $. Therefore we note here that the third condition of a Lie algebra is automatically satisfied by the first two conditions and hence plays no role in finding the Lie closure.

\begin{definition}
The \emph{Lie closure} of a set of matrices, $ \mathcal{A} $, is the intersection of all matrix Lie algebras which contain $ \mathcal{A} $. More simply, this can be described as the smallest matrix Lie algebra which contains $ \mathcal{A} $.
\end{definition}

Once the linear closure of a set of matrices is found, Lie brackets of the linear space's basis elements are calculated: it is sufficient to only work with the basis elements of the linear closure due to the space being linear and the Lie bracket operation being bi-linear. A matrix linear space that is closed under the Lie bracket operation is a Lie algebra. If a Lie bracket is found to be in the existing linear space, the Lie bracket is ignored and another is tried. If a Lie bracket is not in the existing linear space, it is added to the basis. The stop condition is when all Lie brackets of the basis elements are in the linear space. At this point, we have found the Lie closure of the linear space.

\begin{example}
\label{firstLieC}
To find the Lie closure of the matrix set described in Example \ref{firstLC}, we take Lie brackets of the basis elements in the linear closure. We see that

\[ \left( \begin{array}{cc} 1 & 0 \\ 0 & 0 \end{array} \right) \left( \begin{array}{cc} 0 & 1 \\ 0 & 0 \end{array} \right) - \left( \begin{array}{cc} 0 & 1 \\ 0 & 0 \end{array} \right)  \left( \begin{array}{cc} 1 & 0 \\ 0 & 0 \end{array} \right)  = \left( \begin{array}{cc} 0 & 1 \\ 0 & 0 \end{array} \right) \]

\noindent which is in the linear closure so there is no further action to be taken. On the other hand

\[ \left( \begin{array}{cc} 0 & 0 \\ 1 & 1 \end{array} \right) \left( \begin{array}{cc} 1 & 0 \\ 0 & 0 \end{array} \right) - \left( \begin{array}{cc} 1 & 0 \\ 0 & 0 \end{array} \right) \left( \begin{array}{cc} 0 & 0 \\ 1 & 1 \end{array} \right) = \left( \begin{array}{cc} 0 & 0 \\ 1 & 0 \end{array} \right) \]

\noindent which is not in the linear closure so it is added to the basis. Therefore our new space is

\[ \text{span}_{\mathbb{R}} \left\{
\left( \begin{array}{cc} 1 & 0 \\ 0 & 0 \end{array} \right),
\left( \begin{array}{cc} 0 & 1 \\ 0 & 0 \end{array} \right),
\left( \begin{array}{cc} 0 & 0 \\ 1 & 1 \end{array} \right), 
\left( \begin{array}{cc} 0 & 0 \\ 1 & 0 \end{array} \right) \right\} \]

\noindent which is a four dimensional matrix linear space. As this space now spans all $ ( 2 \times 2 ) $ matrices, we know that any Lie bracket in this space will be contained in the space so we now have the Lie closure of the matrix set. 
\end{example}

\begin{example}
\label{HKYLieC}
In Example \ref{HKYlclv}, the linear closure is in the LMM hierarchy \citep{fernandez2015lie} and hence is a Lie algebra so no further computation is necessary to find the Lie closure of that matrix set (so the Lie closure of $ \mathcal{Q}_{HKY'} $ is a four dimensional matrix linear space). The linear version of $ \mathcal{Q}_{HKY'} $ also forms a Lie algebra so the Lie closure of the linear version of $ \mathcal{Q}_{HKY'} $ is equal to the linear version (a four dimensional matrix linear space).
\end{example}

\begin{figure}
  \includegraphics[width=0.75\textwidth]{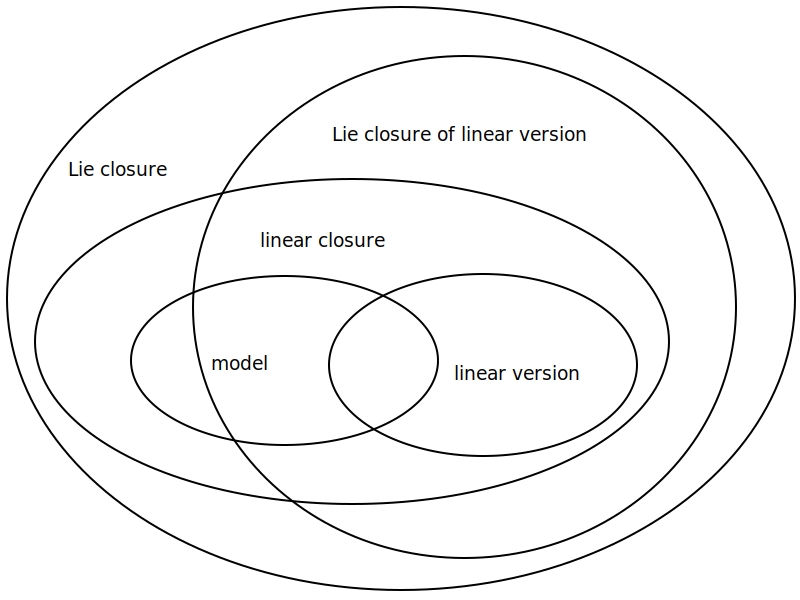}
\caption{ If we begin with a model, the other matrix sets generated by the model will be nested as shown. For example, linear closures being inside Lie closures represents that when the Lie closure and linear closure are not equal then the Lie closure is the larger space of which the linear closure is a linear subspace of. The Lie closure of the linear version does not fit into this diagram easily as it can be both inside or outside of the linear closure, however it will be inside the Lie closure as a simple consequence of Lemma \ref{kitten}. }
\label{fig:1}       
\end{figure}


For the general case, to find the Lie closure of a model, first the linear closure is found and then Algorithm \ref{fitz} is used. An alternative method for finding a Lie algebra associated with a model is to find the Lie closure of its linear version. The nesting of the spaces that can be generated from a model is depicted in Figure \ref{fig:1}. It was found in general practise that Algorithm \ref{fitz} could be computationally expensive to complete: it would ``settle'' on a result in less than a day (i.e. the basis did not increase in size for several thousand iterations) but would have taken months to confirm the result so in these cases we applied Algorithm \ref{confirmationAlg} to check if the resulting space \textit{most likely} formed a Lie algebra or whether there was evidence to suggest that it did not. Algorithm \ref{confirmationAlg} generates Lie brackets in the existing space then tries to solve the resulting matrix in terms of the space's basis. The idea behind it is that it would be very unlikely for the Lie bracket of two matrices which are a linear combination of all the basis elements (whose coefficients are randomly generated rational numbers) to be linear combination of the basis elements if it the space were not a Lie algebra. For the purposes of our study, Algorithm \ref{fitz} produces a lower bound for the size of the Lie closure which is all we need to know it is not going to produce a useful model.


\begin{algorithm}
\SetAlgoLined
 Our input is the basis for the linear space $ L $: $ \{ A_{1}, A_{2}, ... A_{n} \} $\;
 \BlankLine
 $ V = [1,n], $ \#$V$ tracks the dimension of $L $ at certain stages of the algorithm \# \;
 $ k=1 $\;
 \While{$V[k] \not = V[k+1]$}{
 $ i=1 $\;
 \While{$ i \leq V[k+1] $}{
 $ j = V[k] $\;
 \While{$ j \leq V[k+1] $}{
 \eIf{$i < j$}{
  Attempt to solve Lie$ (A_{i}, A_{j}) = a_{1} A_{1} + a_{2} A_{2} + ... + a_{m} A_{m} $ where $m$ is the current dimension of $ L $ and $ a_{l} \in \mathbb{R} $ (note that this essentially solving a linear system of equations with $m$ unknowns)\;
  \eIf{\text{Solution can be found}}
   {None\;}{Lie$(A_{i}, A_{j})$ is appended to $ L $\;}
  }{
  None\;
  } $ j = j+1 $\; } $ i = i+1 $\; }
  Append $V$ with current dimension of $ L $ \;
  $k = k+1$\;
 }
 \bigskip
 \label{fitz}
 \caption{For a given matrix linear space, this algorithm will give the Lie closure.}
\end{algorithm}

\begin{algorithm}
\SetAlgoLined
Our input is a basis for a linear space that we suspect is a Lie algebra $ L $: $ \{ A_{1}, A_{2}, ... A_{n} \} $\;
\BlankLine
$ i=0 $\;
\While{$i<150$}{
$ M_{1} = a_{1} A_{1} + a_{2} A_{2} + ... + a_{n} A_{n} $ and $ M_{2} = b_{1} A_{1} + b_{2} A_{2} + ... + b_{n} A_{n} $ where $ a_{k}, b_{k} $ are random rational numbers between -1000 and 1000\;

Attempt to solve Lie($M_{1}$, $M_{2}$)  $ = c_{1}A_{1} + c_{2}A_{2} + ... + c_{n}A_{n} $\;

\eIf{Solution can be found}{None\;}{Break: our space is not a Lie algebra\;}

$i=i+1$\;
}
If $i$ reaches 150 without a break in the while loop, we conclude that it is highly probable that $L$ is a Lie algebra
\bigskip
\label{confirmationAlg}
\caption{For a given matrix linear space, this algorithm will probabilistically test if the set is a Lie algebra or not. Note that the choice of 150 as the number of iterations was user chosen as a number which was high enough to provide a test of high probability of finding if a space was not a Lie algebra but low enough to ensure fast computation.}
\end{algorithm}

\subsection{Examples of linear closures and linear versions in DNA models} 
~\\

\noindent To better illustrate the different ways of closing matrix sets, and to demonstrate the differences between Lie closures and Lie closures of linear versions, we now further explore finding closures of some popular DNA rate substitution models.

\begin{example}
\label{symLieC}
The symmetric DNA model (GTR \citep{tavare1986} with uniform base distribution) has matrix form 

\[ \mathcal{Q}_{SYM} = \left\{ \left( \begin{array}{cccc} *& a & b & c \\ a &*& d & e \\ b & d &*& f \\ c & e & f &*\\ \end{array} \right): a,b,c,d,e,f \in \mathbb{R} \right\}. \]


Clearly, $ \mathcal{Q}_{SYM} $ is a linear space as all constraints on it are linear, e.g. $ Q_{12} = Q_{21} $. 

We define the symmetric matrix $ S_{ij} $ as the $(4 \times 4)$ matrix with unit entries in positions $(i,j)$ and $(j,i)$, zero entries in all other off-diagonal entries and diagonal entries are set to ensure zero column sum. We also define an anti-symmetric matrix $ T_{ij} $ as the $ (4 \times 4) $ matrix with unit entry in position $ (i,j) $, $ -1 $ in position $ (j,i) $, zeros in all other off-diagonal entires and whose diagonal entries are set to give zero column sum. Clearly the set $ \{ S_{12}, S_{13}, S_{14}, S_{23}, S_{24}, S_{34} \} $ is a basis for $ \mathcal{Q}_{SYM} $. 

Each time a Lie bracket is taken of two symmetric matrices, an anti-symmetric matrix is produced. For example consider the Lie bracket $ [S_{12},S_{13}] $:

\begin{gather*}
\left( \begin{smallmatrix} -1 & \phantom{-}1 & \phantom{-}0 & \phantom{-}0 \\ \phantom{-}1 & -1 & \phantom{-}0 & \phantom{-}0 \\ \phantom{-}0 & \phantom{-}0 & \phantom{-}0 & \phantom{-}0 \\ \phantom{-}0 & \phantom{-}0 & \phantom{-}0 & \phantom{-}0 \end{smallmatrix} \right)
\left( \begin{smallmatrix} -1 & \phantom{-}0 & \phantom{-}1 & \phantom{-}0 \\ \phantom{-}0 & \phantom{-}0 & \phantom{-}0 & \phantom{-}0 \\ \phantom{-}1 & \phantom{-}0 & -1 & \phantom{-}0 \\ \phantom{-}0 & \phantom{-}0 & \phantom{-}0 & \phantom{-}0 \end{smallmatrix} \right)
 -
 \left( \begin{smallmatrix} -1 & \phantom{-}0 & \phantom{-}1 & \phantom{-}0 \\ \phantom{-}0 & \phantom{-}0 & \phantom{-}0 & \phantom{-}0 \\ \phantom{-}1 & \phantom{-}0 & -1 & \phantom{-}0 \\ \phantom{-}0 & \phantom{-}0 & \phantom{-}0 & \phantom{-}0 \end{smallmatrix} \right)
\left( \begin{smallmatrix} -1 & \phantom{-}1 & \phantom{-}0 & \phantom{-}0 \\ \phantom{-}1 & -1 & \phantom{-}0 & \phantom{-}0 \\ \phantom{-}0 & \phantom{-}0 & \phantom{-}0 & \phantom{-}0 \\ \phantom{-}0 & \phantom{-}0 & \phantom{-}0 & \phantom{-}0 \end{smallmatrix} \right)
\\ =
\left( \begin{smallmatrix} \phantom{-}0 & \phantom{-}1 & -1 & \phantom{-}0 \\ -1 & \phantom{-}0 & \phantom{-}1 & \phantom{-}0 \\ \phantom{-}1 & -1 & \phantom{-}0 & \phantom{-}0 \\ \phantom{-}0 & \phantom{-}0 & \phantom{-}0 & \phantom{-}0 \end{smallmatrix} \right) \\ = T_{12} - T_{13} + T_{23}. 
\end{gather*}

For distinct $ a,b,c,d $ in $ \{ 1,2,3,4 \} $, the Lie brackets of the basis elements of $ \mathcal{Q}_{SYM} $ can be summarised as follows:

\begin{align*}
[S_{ab},S_{cd}] =& \phantom{-} 0 \\
[S_{ab},S_{ac}] =& \phantom{-} T_{ab} - T_{ac} + T_{bc}.\\
\end{align*}

\noindent Hence we have

\begin{align*}
[S_{23},S_{24}] =& \phantom{-} T_{23} - T_{24} + T_{34}=  U_{1}\\
[S_{32},S_{34}] =& \phantom{-} T_{32} - T_{34} + T_{24}=  U_{2}\\
[S_{12},S_{14}] =& \phantom{-} T_{12} - T_{14} + T_{24}=  U_{3}\\
[S_{12},S_{13}] =& \phantom{-} T_{12} - T_{13} + T_{23} = U_{4},\\
\end{align*}

\noindent where each $ U_{i} $ is a $4 \times 4$ matrix which is both anti-symmetric and doubly stochastic and has zero entries in all of its $ i^{th} $ row and column. We see that $ U_{4} = U_{1} - U_{2} + U_{3} $ and hence the $ \{ U_{1}, U_{2}, U_{3} \} $ is a linearly independent set which spans the space of doubly stochastic anti-symmetric matrices. The set $ \text{span}_{\mathbb{R}} \{ S_{12}, S_{13}, S_{14}, S_{23}, S_{24}, S_{34}, U_{1}, U_{2}, U_{3} \} $ is the set of doubly stochastic $ 4 \times 4 $ matrices which is known as the doubly stochastic model (DS) and is Model 9.20b in the LMM hierarchy \citep{fernandez2015lie}. It forms a Lie algebra and hence is the Lie closure of SYM.

\end{example}

\begin{example}
\label{GTRLieC}
Consider the GTR model \citep{tavare1986} which has matrix form

\[ \mathcal{Q}_{GTR} = \left\{ \left( \begin{array}{cccc} * & \alpha \pi_{A} & \beta \pi_{A} & \gamma \pi_{A} \\ \alpha \pi_{G} & * & \delta \pi_{G} & \varepsilon \pi_{G} \\ \beta \pi_{C} & \delta \pi_{C} & * & \eta \pi_{C} \\ \gamma \pi_{T} & \varepsilon \pi_{T} & \eta \pi_{T} & *  \end{array} \right): \alpha, \beta, \gamma, \delta, \varepsilon, \eta, \pi_{A}, \pi_{G}, \pi_{C}, \pi_{T} \in \mathbb{R} \right\}. \]

\noindent Its linear version is 
\[ \left\{ \left( \begin{array}{cccc} * & \alpha+ \pi_{A} & \beta+ \pi_{A} & \gamma+ \pi_{A} \\ \alpha+ \pi_{G} & * & \delta+ \pi_{G} & \varepsilon+ \pi_{G} \\ \beta+ \pi_{C} & \delta+ \pi_{C} & * & \eta+ \pi_{C} \\ \gamma+ \pi_{T} & \varepsilon+ \pi_{T} & \eta+ \pi_{T} & *  \end{array} \right): \alpha, \beta, \gamma, \delta, \varepsilon, \eta, \pi_{A}, \pi_{G}, \pi_{C}, \pi_{T} \in \mathbb{R} \right\} \]
\noindent which, although presented using ten free parameters, is a nine dimensional linear space. We define $L_{i}$ to be the matrix generated when each parameter except $i$ is set to be zero, and $i$ is set to be 1. We can then assert that the set $ \{ L_{\alpha}, L_{\beta}, L_{\gamma}, L_{\delta}, L_{\varepsilon}, L_{\eta}, \allowbreak  L_{\pi_{A}},L_{\pi_{G}}, L_{\pi_{C}},L_{\pi_{T}} \} $ is not linearly independent as $ L_{\alpha}+L_{\beta}+L_{\gamma}+L_{\delta}+L_{\varepsilon}+L_{\eta} - L_{\pi_{A}}- L_{\pi_{G}}- L_{\pi_{T}} = L_{\pi_{C}} $.

In its own right, this set of matrices does not form a Lie algebra and therefore the Lie closure of this space is not trivial. The linear version of $\mathcal{Q}_{GTR}$ is not contained in the DS model and hence must be contained in a LMM of a higher dimension than 9. It is not contained in Models 10.12 or 10.34 of the LMM hierarchy \citep{fernandez2015lie} and hence we conclude that the Lie closure of the linear version of $\mathcal{Q}_{GTR}$ must contain the set of $ 4 \times 4 $ matrices which have zero column sum, known as the General Markov Model (GMM) \citep{barry1987}. As GMM is the largest $ 4 \times 4 $ rate matrix set, we conclude that Lie closure of the linear version of $\mathcal{Q}_{GTR}$ cannot be bigger than GMM and is therefore equal to GMM.
\end{example}

\begin{example}
\label{GTRstLieC}
We take the GTR model \citep{tavare1986} and assume that $ \pi_{A} = \pi_{G} $ and $ \pi_{C} = \pi_{T} $. This model has matrix form

\[ \mathcal{Q}_{GTR'} = \left\{ \left( \begin{array}{cccc} * & \alpha \pi_{A} & \beta \pi_{A} & \gamma \pi_{A} \\ \alpha \pi_{A} & * & \delta \pi_{A} & \varepsilon \pi_{A} \\ \beta \pi_{C} & \delta \pi_{C} & * & \eta \pi_{C} \\ \gamma \pi_{C} & \varepsilon \pi_{C} & \eta \pi_{C} & *  \end{array} \right): \alpha, \beta, \gamma, \delta, \varepsilon, \eta, \pi_{A}, \pi_{C} \in \mathbb{R} \right\}. \]

Its linear version has the form
\[ \left\{ \left( \begin{array}{cccc} * & \alpha +\pi_{A} & \beta+ \pi_{A} & \gamma+ \pi_{A} \\ \alpha+ \pi_{A} & * & \delta +\pi_{A} & \varepsilon+ \pi_{A} \\ \beta+ \pi_{C} & \delta +\pi_{C} & * & \eta+ \pi_{C} \\ \gamma+ \pi_{C} & \varepsilon+ \pi_{C} & \eta+ \pi_{C} & *  \end{array} \right): \alpha, \beta, \gamma, \delta, \varepsilon, \eta, \pi_{A}, \pi_{C} \in \mathbb{R} \right\}. \]

We see that $ \mathcal{Q}_{SYM} $ is contained in this set so the Lie closure of the linear version must contain the DS model. We also note that $ L_{\pi_{A}} $ and $ L_{\pi_{C}} $ are not doubly stochastic so the Lie closure of the linear version of $ \mathcal{Q}_{GTR'} $ must have dimension 11 at minimum. As there are no 11 dimensional Lie-Markov models \citep{fernandez2015lie}, the Lie closure of the linear version of $ \mathcal{Q}_{GTR'} $ must be, again, GMM.

We note here that as the Lie closure of the linear version is contained in or equal to the Lie closure of a set, the Lie closures of both $ \mathcal{Q}_{GTR} $ and $ \mathcal{Q}_{GTR'} $ are also GMM.

\end{example}

\begin{example}
\label{TNLieC}
The model proposed by \citet{tamura1993estimation}, often referred to the Tamura Nei model (TN), has the matrix form:

\[ \mathcal{Q}_{TN} = \left\{ \left( \begin{array}{cccc}
* & \pi_{A} \kappa_{1} &  \pi_{A} &  \pi_{A} \\
 \pi_{G} \kappa_{1} & * &  \pi_{G} &  \pi_{G} \\
 \pi_{C} &  \pi_{C} & * &  \pi_{C} \kappa_{2} \\
 \pi_{T} &  \pi_{T} &  \pi_{T} \kappa_{2} & *  
\end{array} \right): \kappa_{1}, \kappa_{2}, \pi_{A}, \pi_{G}, \pi_{C}, \pi_{T} \in \mathbb{R} \right\}. \]

This is an interesting example as both its linear closure and linear version form Lie algebras. Since these models also have purine/pyrimidine symmetries\footnote{We note that the LM models given in \citet{fernandez2015lie} have purine/pyrimidine symmetry which means if we permute nucleotides such that the partitioning of nucleotides into purine/pyrimidine is unchanged, then we obtain a rate matrix that belongs to the same model.}, they are in the LMM hierarchy given in \citet{fernandez2015lie}. The linear version has the form

\[ \left\{ \left( \begin{array}{cccc}
* & \pi_{A} + \kappa_{1} &  \pi_{A} &  \pi_{A} \\
 \pi_{G}+ \kappa_{1} & * &  \pi_{G} &  \pi_{G} \\
 \pi_{C} &  \pi_{C} & * &  \pi_{C}+ \kappa_{2} \\
 \pi_{T} &  \pi_{T} &  \pi_{T}+ \kappa_{2} & *  
\end{array} \right): \kappa_{1}, \kappa_{2}, \pi_{A}, \pi_{G}, \pi_{C}, \pi_{T} \in \mathbb{R} \right\} \]

\noindent which is Model 6.8a of the LMMs \citep{fernandez2015lie}. The linear closure on the other hand is

\[ \left\{ \left( \begin{array}{cccc}
* & \alpha &  \pi_{A} &  \pi_{A} \\
 \beta & * &  \pi_{G} &  \pi_{G} \\
 \pi_{C} &  \pi_{C} & * &  \gamma \\
 \pi_{T} &  \pi_{T} &  \delta & *  
\end{array} \right): \alpha, \beta, \gamma, \delta, \pi_{A}, \pi_{G}, \pi_{C}, \pi_{T} \in \mathbb{R} \right\} \]

\noindent which is Model 8.8 of the LMMs \citep{fernandez2015lie}.

\end{example}

In summary we have: 
\begin{res}
The Lie closure of $\mathcal{Q}_{SYM}$ is the DS model. The Lie closures, and Lie closures of the linear versions, of both GTR and GTR' are GMM. The Lie closure of TN is Model 8.8 of the LMMs and the Lie closure of the linear version of TN is Model 6.8a of the LMMs.
\end{res}

\begin{proof}
As established in Examples \ref{symLieC}, \ref{GTRLieC}, \ref{GTRstLieC} and \ref{TNLieC} above.
\qed
\end{proof}

\subsection{Incorporating the $ \omega $ parameter into Lie closures of codon models} 
The use of linear versions (which leads to use of Lie closures of linear versions) changes the $ \omega $ parameter from a multiplicative operation to an additive one (for examples of the unique rates present in Markov models of interest, see Table \ref{tab:2}). Adding scalar multiples of the matrix $G$ to an existing rate matrix would result in undesirable consequences, for example, non-zero entries where multiple nucleotide substitutions are required. We are therefore required to define a new matrix, $G^*$, with the parameter $\omega$ being the coefficient of this matrix in linear versions. In off-diagonal entries, the matrix $G^*$ is defined to have unit entries for the entries representing non-synonymous mutations which require only one nucleotide mutation and are not to or from stop codons; and zero entries everywhere else. Its diagonal entries are chosen to give zero column sum. Additionally, because we are not multiplying $ Q_{triplet} $ (\ref{Qtrip}) by $G$, we are required to add the extra constraint on $ Q_{triplet} $ of zero values for entries that represent mutations to or from stop codons. When the linear version is found for a MG-style codon model, $G^*$ is automatically in the basis and when the matrix set is written as a linear combination of its basis elements, $G^*$ would have the coefficient $\omega$. This accounts for all ``$ + \omega $'' terms of off diagonal matrix entries (for examples of matrix entries in linear versions of MG-style codon models, see Table \ref{tab:2}).

Defining an $ \omega $ parameter for the linear closure (and hence Lie closure) of an MG-style codon model case is less clear. When the linear closure of the codon model is found, for example, of K2ST-MG we start with parameters $ \{ a, b, \omega \}$ (which would result in matrix entries $\{ a, b, a \omega, b \omega \} $) and the linear closure has parameters $ \{ a, b, a \omega, b \omega \} \rightarrow \{ c_{1}, c_{2}, c_{3}, c_{4} \} $ (\emph{i.e.} there are now 4 independent parameters) which means that there is no longer a clear $ \omega $ parameter. Like the linear version, it would seem logical for $ \omega $ to be the coefficient of the $ G^* $ matrix. Therefore in practice, the basis for the linear closure should be defined in a way to include $ G^* $. This still leaves the question of how $ \omega $ itself should be calculated.

One possible way to calculate $ \omega $ is $ \frac{1}{2} (\frac{c_{3}}{c_{1}} + \frac{c_{4}}{c_{2}}) $; an average of the non-synonymous/synonymous rate ratios. Another method proposes that $ \omega_{1} = \frac{c_{3}}{c_{1}} $, $ \omega_{2} = \frac{c_{4}}{c_{2}} $ and hence $ \omega = \omega_{1} \frac{c_{1}}{c_{1} + c_{2}} + \omega_{2} \frac{c_{2}}{c_{1} + c_{2}} $; a weighted average where the weights are the frequencies of the types of substitutions. A third method would be to consider the geometric average between $\frac{c_{3}}{c_{1}}$ and $\frac{c_{2}}{c_{3}}$ so we would obtain $\omega = \sqrt{\frac{c_{3}c_{4}}{c_{1}c_{2}}}$. It is currently an open question to how $ \omega $ is to be calculated or interpreted; especially as the situation is more complicated in MG-style codon models whose linear closures have more than 4 parameters.


\begin{table}
\caption{Interesting cases of Markov models, codon and DNA, with their unique off-diagonal entries and the number of free parameters listed. Note that $ i \in \{ \text{A, G, C, T} \} $. For the linear versions and linear closures, the $\pi_i$ should simply be thought of as free parameters since they are no longer proportional to the DNA equilibrium frequencies. Finding the codon model of a linear version of HKY then finding the linear version of that model is equivalent to finding the linear version of HKY-MG; this also applies to linear closures. }

\label{tab:2}
\begin{tabular}{l|ll}
\hline\noalign{\smallskip}
Model & unique off-diagonal matrix entries & \#parameters \\
\noalign{\smallskip}\hline\noalign{\smallskip}
JC  & $\alpha$ & 1 \\
JC-MG & $\alpha$, $\alpha\omega$ & 2 \\
JC-MG: linear version & $\alpha$, $\alpha + \omega$ & 2 \\
JC-MG: linear closure & $\alpha$, $\alpha'$ & 2 \\
\noalign{\smallskip}\hline\noalign{\smallskip}
K2ST  & $ \alpha $, $\beta$ & 2 \\
K2ST-MG & $ \alpha $, $\alpha\omega$, $\beta$, $\beta\omega$ & 3 \\
K2ST-MG: linear version & $ \alpha $, $\alpha + \omega$, $\beta$, $\beta + \omega$ & 3 \\
K2ST-MG: linear closure & $ \alpha $, $\alpha'$, $\beta'$, $\beta$ & 4 \\
\noalign{\smallskip}\hline\noalign{\smallskip}
HKY  &  $ \pi_{i} $, $ \pi_{i}\kappa $ & 5 \\
HKY:  linear version & $ \pi_{i} $, $ \pi_{i} + \kappa $ & 5 \\
HKY:  linear closure & $ \pi_{i} $, $ \pi_{i}' $ & 10 \\
HKY-MG & $ \pi_{i} $, $ \pi_{i}\kappa$, $\pi_{i}\omega $, $ \pi_{i}\kappa\omega $ & 6 \\
HKY-MG: linear version & $ \pi_{i} $, $ \pi_{i}+\kappa$, $\pi_{i}+\omega $, $ \pi_{i}+\kappa+\omega $ & 6 \\
HKY-MG: linear closure & $ \pi_{i} $, $\pi_{i}'$, $\pi_{i}''$, $\pi_{i}'''$ & 16 \\
\noalign{\smallskip}\hline
\end{tabular}

\end{table}

\subsection{Lie closures of codon models}

In our analysis of codon models, first the codon model was defined in the way we have discussed in Section \ref{sec:1}. Recall that under this model structure the rates of change between codons under JC-MG and K2ST-MG are as follows:

\[ Q_{\text{JC-MG}} = \begin{cases}
0 & \text{mutations that are to or from stop codons} \\
0 & \text{multiple nucleotide substitutions required} \\
\alpha & \text{synonymous substitution} \\
\alpha\omega & \text{non-synonymous substitution} \\
\end{cases}\]

\[ Q_{\text{K2ST-MG}} = \begin{cases}
0 & \text{mutations that are to or from stop codons} \\
0 & \text{multiple nucleotide substitutions required} \\
\alpha & \text{synonymous transition} \\
\alpha\omega & \text{non-synonymous transition} \\
\beta & \text{synonymous transversion} \\
\beta\omega & \text{non-synonymous transversion} \\
\end{cases}\]

\noindent where $\alpha$ and $\beta$ represent substitution behaviour at a DNA level.

The linear closure was found for both JC-MG and K2ST-MG, the linear version of K2ST-MG was found, and then Algorithm \ref{fitz} was applied to these linear spaces. It should be noted here that the linear closure and linear version of JC-MG are the same linear space only with different bases and hence have the same Lie closure. This is not the case for K2ST-MG, for this codon model the linear closure and linear version are different linear spaces; this is because the model has non-linear constraints. We waited for the algorithm to ``settle'' then, as the algorithm would have taken months to terminate, used Algorithm \ref{confirmationAlg} to confirm that the resulting spaces were Lie algebras.



We found the dimensions of both the Lie closure of K2ST-MG and the Lie closure of the linear version of K2ST-MG are 2106 (curiously, these were found to be the same linear space which is a topic for further exploration). The dimension of the Lie closure of JC-MG was 1996. It is therefore clear that such models are far too big to be of practical use in phylogenetic applications. In order for the Lie closure or the Lie closure of the linear version to be smaller, the starting model would have to be simpler but the only way we can make a codon model that is simpler than JC-MG is to set $ \omega $ to a constant value which would ruin the whole point of the exercise as we are trying to reduce mis-estimation of $ \omega $.

\subsection{Further analysis: partial Lie closures}
It was thought, given that finding a full Lie closure of a MG-style codon model was not practical, that we could instead create a \emph{partial Lie closure}; that is to begin to close the Lie algebra but not completely do so. We now define more precisely what we mean by a partial Lie closure.

In Algorithm \ref{fitz} above, we can see that any element added to the basis $ L $ can be represented as a Lie bracket of the original $n$ matrices from the linear closure. We define the \emph{generation} of an element of $L$ as the number of Lie brackets necessary to build that element from the elements of the linear closure plus one. For example, we would say that $ B = [A_{1},[[A_{2},A_{3}],A_{1}]] $ (where $ A_{1}, A_{2}, A_{3} $ are in the linear closure of the original matrix set) would belong to generation $4$ as there are $3$ Lie bracket operations required to build this element from the elements of the linear closure. When building a partial Lie closure, we will calculate elements up to a fixed generation. For example, if one was interested in a Lie closure up to generation 4 then first generation 1 elements would be calculated followed by generations 2, 3 and 4. This process is really the same as the algorithm for finding the Lie closure (described in Algorithm \ref{fitz}) apart from the stop condition and the order in which Lie brackets are calculated (and possibly appended to $ \mathcal{L} $).

It was hoped that a model that was partially Lie closed would have similar enough properties to Lie algebras that the mis-estimation of  $ \omega $ could be reduced. We tried to find a partial closure of the JC-MG model. Unfortunately, problems arose regarding ``stochasticity.''

We say a zero column sum matrix is stochastic when it is a rate matrix i.e. a matrix is stochastic when its off diagonal entries are non-negative. This is a requirement of rate matrices as it does not make sense to have a negative rate of one state changing to another. Sometimes given a matrix linear space, constraints must be put on the basis coefficients in order to achieve stochasticity. For example, for the linear space of matrices



\[ \mathcal{A} = \text{span}_{\mathbb{R}} \left\{ A_{1} = \left( \begin{array}{ccc} -2 & 1 & 1 \\ 1 & -2 & 1 \\ 1 & 1 & -2 \end{array} \right), A_{2} =  \left( \begin{array}{ccc} 0 & -1 & -1 \\ -1 & 0 & 1 \\ 1 & 1 & 0 \end{array} \right) \right\} \]


\noindent with a typical element $ a_{1}A_{1} + a_{2}A_{2}: a_{1},a_{2} \in \mathbb{R} $, we must place constraints on $ a_{1} $ and $ a_{2} $ in order for matrices in $ \mathcal{A} $ to be stochastic. One possible set of constraints is $ a_{1} \geq 0 $ and $ a_{1} \geq |a_{2}| $. Sometimes however, there are no constraints that will ensure non-trivial stochasticity, for example consider the set

\[ \mathcal{B} = \text{span}_{\mathbb{R}} \left\{ B_{1} = \left( \begin{array}{ccc} -2 & 1 & 1 \\ 1 & -2 & -1 \\ 1 & 1 & 0 \end{array} \right), B_{2} =  \left( \begin{array}{ccc} -2 & 1 & -2 \\ 1 & -2 & 1 \\ 1 & 1 & 1 \end{array} \right) \right\} \]

\noindent with the typical element $ b_{1}B_{1} + b_{2}B_{2}: b_{1},b_{2} \in \mathbb{R} $. We see that the only way an element of $ \mathcal{B} $ can be stochastic is if we set $ b_{1} = b_{2} = 0 $.

It was found that any non-trivial partial Lie closure (\emph{i.e.} a partial Lie closure which is bigger than the linear closure where there can be non-zero coefficients for the basis elements that are not in the linear closure) of the JC-MG codon model with dimension of less than 227 (this was finding the partial Lie closure up to generation $10$) violated stochasticity. This means that for a non-trivial partial Lie closure to be stochastic, we would need a dimension $ \geq 227 $ but such a space is still too big to be practical.

\section{Toy model: an interesting case of symmetries} 
\label{sec:3}
It is interesting that the Lie closure of a codon model which began with a linear space with a dimension of 2 could have a Lie closure whose dimension is so large. Studying this further has proven to be difficult given the computational difficulty of the problem. A toy model was created in an attempt to better understand the features that could lead to the Lie closure of a linear space being so large.

We assumed that the codon length was $3$. We then assumed that the number of states is $2$ (R and Y) instead of $4$ (A, G, C and T). The resulting codon model is $ (8 \times 8) $. Like in the MG-style codon models, it was assumed that there cannot be two changes happening on the same codon at once so, for example, the rate of $ RRR \rightarrow RYY = 0 $. We defined our basis model as

\[ Q_{triplet} = Q_{2} \otimes I \otimes I + I \otimes Q_{2} \otimes I + I \otimes I \otimes Q_{1} \]

\noindent where 

\[ Q_{1} = \left( \begin{array}{rr} -a & a \\ a & -a \end{array} \right) \text{ and } Q_{2} = \left( \begin{array}{rr} -b & b \\ b & -b \end{array} \right) .\]

\noindent This results in 

\[ Q_{triplet} = \left( \begin{array}{cccccccc}
* & a & b & 0 & b & 0 & 0 & 0 \\
a & * & 0 & b & 0 & b & 0 & 0 \\
b & 0 & * & a & 0 & 0 & b & 0 \\
0 & b & a & * & 0 & 0 & 0 & b \\
b & 0 & 0 & 0 & * & a & b & 0 \\
0 & b & 0 & 0 & a & * & 0 & b \\
0 & 0 & b & 0 & b & 0 & * & a \\
0 & 0 & 0 & b & 0 & b & a & * \\
\end{array} \right). \]

This matrix is equivalent to a full codon model where a synonymous change is when the third codon position mutates to another nucleotide and a non-synonymous change is when the first or second codon position mutates to another nucleotide. As it currently stands, $ Q_{triplet} $ forms an abelian Lie algebra so its Lie closure would have dimension $2$. What we want to test now is if we make minor adjustments to $ Q_{triplet} $, what will happen to the size of the Lie closure?

If matrix entry $ (7,5) $ of $ Q_{triplet} $ is changed from $b$ to $a$, then the dimension for the Lie closure is $25$. When similar changes were made to $ Q_{triplet} $ (swapping $b$ values to $a$ values and vice versa), the dimensions of the Lie closures ranged from $5$ to $56$. (Note that the maximum possible size for a Lie closure of a model of this form is $ 8 \times 8 - 8 = 56 $.) There was an apparent trend that when the adjusted $ Q_{triplet} $ was still symmetrical after being altered (i.e. if we changed the matrix entry $(7,5)$ from $b$ to $a$ then we also changed the entry $(5,7)$ from $b$ to $a$) the Lie closure tended to be smaller but such symmetries were not sufficient to obtain a Lie closure of less than $ 5 $.

For a fixed number of changes, we found there was great variety in the size of the resulting Lie closure. Table \ref{tab:3} gives details on the sizes of the Lie closures generated after making 4 symmetric changes (2 pairs of changes) in 10 different ways and making 2 asymmetric changes in 10 different ways. For more details on the range of Lie closure sizes after 10 various ways of making a particular number of changes, see Table \ref{LCsizes}. We notice that the more changes we make, generally the larger the resulting Lie closure is. It can also be seen that, for the 10 possibilities we tried, having four or more asymmetric changes resulted in a Lie closure size which is as high as possible.

%


\begin{table}
\caption{This ``stem and leaf'' plot shows displays sizes of Lie closures after making various symmetric and asymmetric changes to $ Q_{triplet} $ (for the symmetric case, there were 4 changes made and for the asymmetric case, there were 2 changes made). Key: symmetric $ 1|3| = 31 $, asymmetric $ |3|1 = 31 $.}

\label{tab:3}
\begin{tabular}{r|c|l}
\hline\noalign{\smallskip}
symmetric  &  &  asymmetric \\
\noalign{\smallskip}\hline\noalign{\smallskip}
  5  5 & 0 &     \\
    7  0  0 & 1 & 3     \\
  & 2 & 5  8   \\
  9  7  2  2 & 3 & 3  7  7  \\
   9 & 4 & 9  9  9  9  9   \\
\noalign{\smallskip}\hline
\end{tabular}

\end{table}

\begin{table}
\caption{The observed size range of Lie closures of matrices being $ Q_{triplet} $ with a particular number of modifications in 10 different ways.}
\label{LCsizes}
\begin{tabular}{c|cc}
     $n$ & $2n$ symmetric changes & $n$ asymmetric changes  \\ \hline
     \begin{tabular} c \\ 1\\ 2 \\ 3 \\ 4\\ \end{tabular} & 
     \begin{tabular}{cc} min & max \\ \hline  5&17 \\ 5 &37 \\ 18&37\\25&49 \end{tabular} &
     \begin{tabular}{cc} min & max \\ \hline  13&25 \\ 13 &49 \\ 49&56\\56&56 \end{tabular} 

\end{tabular}
\end{table}



This is interesting as it shows that as soon as the model deviates from being $ Q_{triplet} $ the Lie closure no longer has a simple answer. The linear closures of JC-MG and K2ST-MG are far from being as simple as the $ (64 \times 64) $ $ Q_{triplet} $ matrix so it is not surprising that the Lie closures are so large.

\section{Discussion} 
\label{sec:4}

Our initial aim in this work was to find a multiplicatively closed codon model which incorporated $ \omega $ as a parameter. We have shown that there is no practical way to do this. 

This negative result is interesting in a mathematical context. In the case of the JC-MG codon model, it was surprising that a model which began with as few as 2 parameters would have a Lie closure of 1996 parameters. Our investigation of the toy model demonstrated that this was intrinsic to the problem: if the starting matrix set deviates too far from being perfectly symmetrical, then the Lie closure tends to be close to as large as possible.


From here, the perceived way forward is to conduct further analysis of linear closures and linear versions of codon models. These do not violate stochasticity and are reasonable in size. In our case of analysing the JC-MG model, the linear closure (which is the same as the linear version, only a different basis) is trivial due to JC-MG only having two parameters to begin with. But when the underlying DNA rate substitution process has more parameters, for example HKY, then the linear closure is not trivial and the setup of the linear version is quite different to the original. It has not yet been tested to see if linear closures of codon models mis-estimate $ \omega $ as much as the models themselves but it is possible that this could help as previous exploration \citep{bodie2011effect,sumner2011general} found that, in DNA models, parameters are mis-estimated less in DNA models which form linear spaces.

This approach also brings about the opportunity to further study an additive $ \omega $. As previously discussed, it is not immediately clear how this parameter should be defined in the model building process. Making this parameter additive introduces an opportunity for stochasticity violation in a model which is a mathematical problem yet to be explored. How to interpret the parameter biologically when it is added instead of multiplied is also an open problem which opens up potential research topics.

~\\

\begin{acknowledgements}
We thank Andrey Bytsko for pointing out an error in an early draft regarding computation of the linear closures. We also thank the anonymous reviewers for their thorough reading of the manuscript and insightful comments that have led to a substantially improved article.
\end{acknowledgements}

%
%

\bibliographystyle{spbasic}      
\bibliography{manuscript.bib}   


\end{document}